# Protecting Anti-virus Programs From Viral Attacks

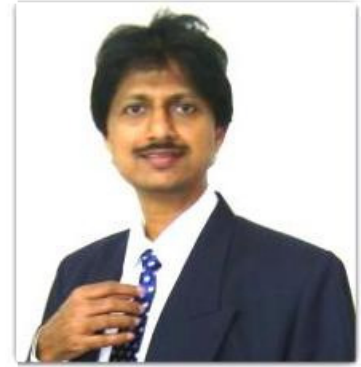

**By- Umakant Mishra, Bangalore, India**
umakant@trizsite.tk, http://umakant.trizsite.tk

**Contents**

## 1. Introduction

The war between virus creators and anti-virus developers started since the birth of the earliest viruses in eighties. Viruses used various stealth techniques such as encryption, polymorphism, metamorphism, anti-emulation and anti-heuristic techniques to escape all types of virus scanning. Besides some viruses took offensive measures to attack the anti-viruses. As they found that an anti-virus program is their biggest enemy they came up with the idea to screw the anti-virus program and paralyze the functions of the anti-virus system. Once the anti-virus is compromised there will be no trouble for the virus to grow and the computer can be a safe have for virus proliferation.

It may be useful to mention that viruses are present in almost every system, which is in contact with external systems in LAN or Internet, even if an anti-virus is installed. If the anti-virus is not reliable or not updated or compromised in some way then it fails to provide the necessary protection to the system and allows the viruses to breed and grow safely in the system. Besides even the best anti-viruses are not perfect and allow many false positives and false negatives.

## 2. Weak Points of an Anti-Virus Program

History shows that every anti-virus program including the popular names like McAfee, Symantec, TrendMicro, VBA32, Panda, PC Tools, CA eTrust, ZoneAlarm, AVG, BitDefender, Avast!, Sohost, Kaspersky etc. have been attacked by different viruses at different times. While an anti-virus is supposed to protect the client's machine from viruses, it has been quite challenging for the anti-virus to protect itself from viruses. Let's mention the reasons why the anti-viruses sometimes get defeated and paralyzed by the viruses.

⇨ Insufficient testing- as the current day anti-virus programs are becoming more and more complex, it has become extremely difficult to make a thorough testing of the product before releasing to the market. This situation leaves some bugs and loopholes in the anti-virus product, such as buffer overflow, heap overflow, integer overflow etc. which are intentionally exploited by the malware programmers.

⇨ Some viruses (rootkits) replace the operating system files with malware substitutes and fool the anti-virus programs by running their own code in the behind.

⇨ Attacking AV files - some viruses attack anti-viruses by replacing their core executables, or altering virus signature databases. Some viruses attack the integrity database to alter the checksums or hash functions in the database. Some viruses try to alter the AV state database and change the status of infected files as "virus free".

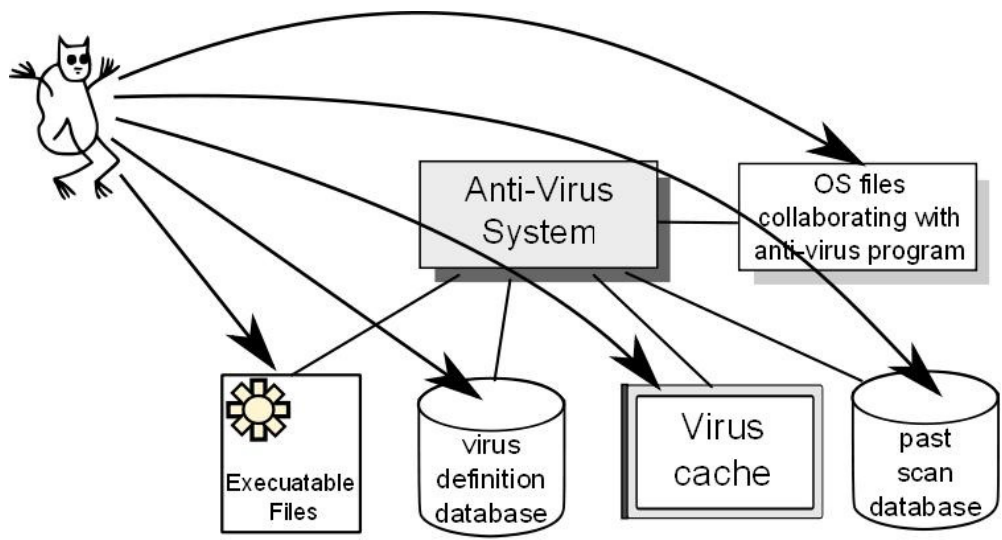

Attacking Different Components of an Anti-Virus system

- ⇨ Some other viruses provide highly compressed bait files to the virus scanner to keep the scanner busy in decompressing. Some viruses try to attack vulnerable ActiveX controls by passing very long strings as parameters.
- ⇨ The last but not the least our overwhelming faith on the anti-virus system often makes us ignore the hygienic habits and deal carelessly with suspicious incidents which leads to a higher possibility of infection. Let's analyze the reasons why do the anti-viruses are targeted by the viruses.

We have already discussed various types of attacks on the anti-virus systems in the previous article ("How do Viruses Attack Anti-Virus Programs"). In this article, we will mainly focus on the remedial measures to protect the anti-virus programs.

## 3. How to Prevent Attacks on Anti-Virus Programs

It is necessary for an anti-virus to protect its files from being attacked by any malware. As it is just a fight between malware and anti-malware the anti-malware has to detect and destroy the malware before its own files are detected and destroyed by the malware.

- ✓ Auditing anti-virus software engine – as we saw above one of the reasons of anti-virus vulnerability is the bugs in their programs. Hence it is necessary to audit the source code of the anti-virus software before finally releasing to the market. There are some ready-made auditing tools available in the market, like FlawFinder, RATS, ITS4, SPLINT, CodeScan, Coverity etc. Reverse engineering is another method of auditing to look for potential vulnerabilities.
- 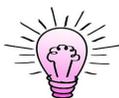Fuzzing is another technique used for testing the loopholes in the anti-virus. The fuzzer may create random files of various types and monitor the behavior of the antivirus software. The anti-virus vendors should fuzz their software before releasing them.
- ✓ In order to detect whether the anti-virus installed on the computer is infected, it is often useful to scan the computer online through an anti-virus service provider. As the code coming from a different source it may be able to detect the infections in the anti-virus system.

### INVENTION-1: Using polymorphic protection method to prevent reverse engineering

Sometimes the virus writers reverse engineer the protection mechanism in order to enable their viruses to escape the available detection mechanism.

Patent 5684875 (invented by Hans Ellenberger, Nov 1997) suggests a polymorphic protection method which installs only a few selected algorithms each time the anti-virus is installed. In this case even if a virus writer reverse engineers the detection method he can view only limited number of algorithms which are installed on that machine. Even if the new viruses include the counter measures for those algorithms they cannot escape for long as the same will be detected by another installation with a different set of algorithms.

### INVENTION-2: Obfuscating anti-virus program files to protect anti-virus programs from virus attack

There is a need for improved method for protecting anti-malware from malware attacks. Patent 7640583 (invented by Marinescu et al., assignee- Microsoft Corporation. Dec 2009) suggests to use all those techniques to hide the anti-malware program which are generally used by a malware to hide its files.

The invention obfuscates or hides the anti-malware program files. The obfuscation may include changing of the identification or file name of the item, changing of the signature of the item and/or changing of the size of the item. (Principle-35: Parameter change). Alternatively, the anti-malware file may be randomly renamed performing polymorphism on the anti-malware file to alter the size and signature of the anti-malware file (Principle-36: State change).

These obfuscation techniques make it difficult for the malware to locate the anti-malware files in order to accomplish the desired task. Malware that attempts to overcome this protection technique will likely include or use a detection engine. But detection engines are large in size and can easily produce identifiable signatures. Thus malware that includes detection engine will be easily caught by existing detection techniques like signature scanning.

### INVENTION-3: Escaping virus interference by scanning the changed sectors of the hard disk

An anti-virus generally mark scanned files as "already scanned" to prevent them from subsequent scanning. But there is a possibility that a virus may change the status of an infected file as "already scanned" thereby causing the scanner to skip that file.

Patent 7401361 (System and method for reducing virus scan time, invented by Freeman, et al., Assignee- Lenovo (Singapore) Pte. Ltd., Jul. 2008) advises to mark only the changed sectors in a hard disk and scan all the changed sectors.

According to this method a protected archive bit is maintained for each sector on the hard disk. When a sector on the hard disk is altered, the sector is marked. This information is stored in a secured location which is shared only with an authorized software like the anti-virus program. When a malware changes the status of a file as "already scanned" and infects the file, the file is still scanned because the changes are marked in the sectors of the hard disk.

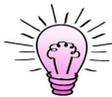
A virus may modify the actual MBR and create a façade MBR at a location other than the actual location of the MBR and fill the façade with the content of the original MBR. When an anti-virus checks the MBR to determine its integrity the operation is intercepted and performed on the façade instead of the actual modified MBR in order to hide the infection. Patent 6802028 (Computer virus detection and removal, invented by Ruff et al., assignee PowerQuest Corporation, Oct 2004) avoids such situations by using a separate BIOS called "trusted BIOS". If any inconsistency is detected between the standard BIOS and trusted BIOS, then the method relies on the trusted BIOS and removes the infected BIOS and reconstructs boot sectors and MBRs.

- ✓ Fixing ActiveX problems- Sometimes the ActiveX controls used by anti-viruses have design errors which include insecure methods. The attackers may pass very long strings as parameters to the vulnerable ActiveX controls in order to cause a memory corruption. This type of problems are treated by auditing the Active-X controls properly before releasing them to the public.

- ✓ Timeout mechanism- A virus may produce highly compressed versions of large files intentionally to put the anti-virus in trouble. As the anti-virus has to decompress the data and executable files before processing them, the anti-virus will use excessive amount of memory making the computer slow or leading to Denial of Service. To avoid such situations some anti-viruses use a time-out mechanism to exit or abort a prolonged scanning operation.

### INVENTION-4: Providing break points in protracted scanning operations

There is a need to break a protracted scanning that exceeds beyond a predefined time. But the situation leads to the following contradiction.

> **Contradiction**
>
> *If a protracted scanning is not timed out, then the anti-virus system unnecessarily uses system resources and impacts system performance. But timing out a protracted scanning may inappropriately terminate a scanning operation in a slow or overloaded computer thereby compromising vulnerabilities. We want to terminate an unduly prolonged scanning but don't want to terminate a prolonged scanning in a genuinely slow or stressed system.*

Patent 6968461 (invented by Lucas, et al. assignee Networks Associates Technology, Nov 2005) provides a method of triggering break points during a scanning operation without exposing the system to vulnerabilities. The method provides an additional degree of sophistication by comparing the size of the computer file being scanned. If the size of a file is large, then it is not early terminated as it may legitimately require a large amount of data to be processed during the scanning operation. (Principle-16: Partial or excessive action, as it is not possible to exactly measure the size of the data processing required).

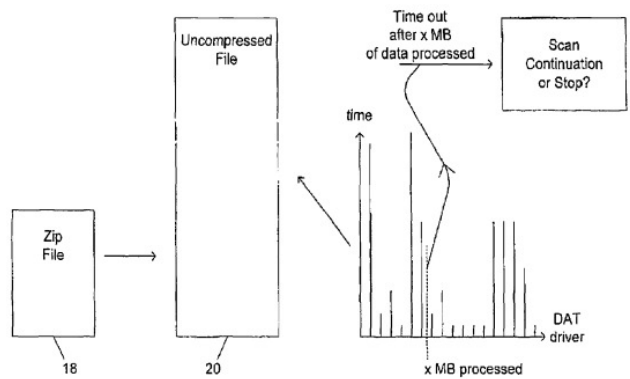

The scanning method first calculates a measurement value based on the amount of data processing performed during the scanning operation and then triggers a break if the said measurement value exceeds a threshold value (Principle-9: Prior Counteraction).

- ✓ Check for the rootkits first before checking viruses. This is because presence of a rootkit can hide other types of viruses/malware. If a rootkit is fooling the anti-virus then any amount of scanning will not help detection of the hidden virus.

- ✓ The anti-virus has to use complex encryption mechanism to protect its virus definition files from being altered or stolen.

## 4. Using TRIZ Inventive Standards to solve this problem

TRIZ Inventive Standards provides some ready made methods to remove the harmful effects of an element on another element of a system. Particularly Inventive Standards of Group-1.2. Decomposition of SFMs are useful for this purpose.

Standard 1.2.1- when there is a harmful interaction between two substances (S1 and S2) introduce another substance (S3) to eliminate this harmful interaction.

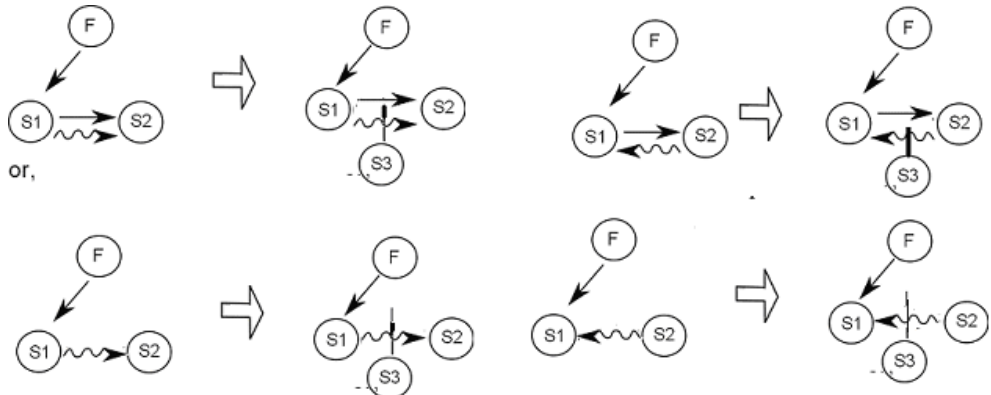
Standard 1.2.1 harmful effect is eleminated

Standard 1.2.2- Introduce a modification of the substances either S1 or S2. While Standard 1.2.1 intoduces a new substance S3 from outside the system, Standard 1.2.2 introduces a substance S3 that is modified or obtained from the existing substances S1 or S2.

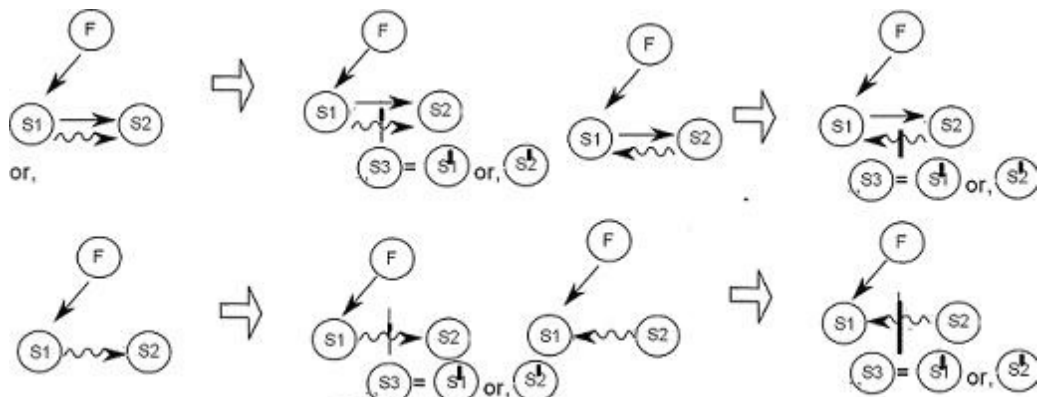
Standard 1.2.2 harmful effect is eleminated

Standard 1.2.3 – Introduce a new substance (S3) that will absorb the harmful effects of either S1 or S2. This solution is different from the former solutions as here the substance S3 absorbs or draws off the harmful effects of the field.

Standard 1.2.4- Introduce a new field (F2) that will neutralize the harmful effects of either S1 or S2. While the previous solutions introduce new substances or modified substance, this method solves the problem by introducing a new field.

Standard 1.2.5- Removing negative effects by removing "ferromagnetic properties" and using physical effects. (This solution has to be suitably modified for software problems).

Some of the solution using the above inventive standards are as follows. We will discuss more solutions using Inventive Standards later in a separate article.

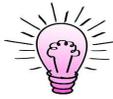 Using polymorphism to protect anti-virus files and codes from the vision of viruses. The anti-virus files may continuously change their names, size, location and other parameters to avoid detection by intelligent viruses.

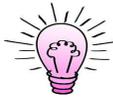 Using a special hardware to store the anti-virus so that the virus cannot make any changes to the anti-virus files or parameters.

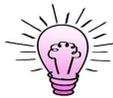 Using a different Operating system for scanning the virus so that the virus of the host operating system will have no impact on the scanning software.

7. Umakant Mishra, "An Introduction to Virus Scanners", http://papers.ssrn.com/sol3/papers.cfm?abstract_id=1916673

8. Umakant Mishra, "Improving Speed of Virus Scanning- Applying TRIZ to Improve Anti-Virus Programs", http://papers.ssrn.com/sol3/papers.cfm?abstract_id=1980638

9. Feng Xue, "Attacking Antivirus", http://www.blackhat.com/presentations/bh-europe-08/Feng-Xue/Whitepaper/bh-eu-08-xue-WP.pdf

10. Peter Szor, The Art of Computer Virus Research and Defense, http://computervirus.uw.hu/index.html

11. Raghunathan Srinivasan, "Protecting Anti-Virus Software under Viral Attacks", http://citeseerx.ist.psu.edu/viewdoc/download?doi=10.1.1.93.796&rep=rep1&type=pdf

12. Vladimir Petrov (edited), "System of Standard Inventive Solutions"

13. US Patent 5684875, "Method and apparatus for detecting a computer virus on a computer", Inventor- Ellenberger, Nov 1997.

14. US Patent 6802028, "Computer virus detection and removal", Inventor- Ruff et al., assignee PowerQuest Corporation, Oct 2004

15. US Patent 6968461, "Providing break points in a malware scanning operation", Inventor- Lucas, et al. assignee Networks Associates Technology, Nov 2005

16. US Patent 7401361, "System and method for reducing virus scan time", Inventor- Freeman, et al., Assignee- Lenovo (Singapore) Pte. Ltd., Jul 2008.

17. US Patent 7640583, "Method and system for protecting anti-malware programs", Inventor- Marinescu et al., assignee Microsoft Corporation, Dec 2009

18. US Patent and Trademark Office (USPTO) site, http://www.uspto.gov/

**Protecting Anti-Virus from Viral Attacks, by Umakant Mishra**     http://umakant.trizsite.tk